\documentclass[letterpaper,aps,prd,preprint,showpacs,nofootinbib,superscriptaddress]{revtex4-1}
\usepackage{amsmath,amssymb,graphicx}
\usepackage[stable]{footmisc}
\usepackage{slashed}

\pdfoutput=1

\newcommand{\nc}{\newcommand}
\nc{\postscript}[2] 
{\setlength{\epsfxsize}{#2\hsize}\centerline{\epsfbox{#1}}}
\nc{\non}{\nonumber}
\nc{\hc}{\hbox {h.c.}} \nc{\re}{\hbox {Re}} 
\nc{\mev}{\hbox {MeV}} \nc{\gev}{\;\hbox {GeV}} \nc{\tev}{\;\hbox {TeV}}
\def\lsim{\mathrel{\raise.3ex\hbox{$<$\kern-.75em\lower1ex\hbox{$\sim$}}}}
\def\gsim{\mathrel{\raise.3ex\hbox{$>$\kern-.75em\lower1ex\hbox{$\sim$}}}}

\nc{\etal}{{\it et al.}}
\nc{\Lsp}{\;\;\;\;\;\;\;\;\;\;}  \nc{\LLLsp}{\lspace \lspace}
\nc{\lsp}{\;\;\;\;\;\;}
\nc{\spac}{\;\;\;}
\nc{\noi}{\noindent}
\nc{\beq}{\begin{equation}}   \nc{\eeq}{\end{equation}}
\nc{\bea}{\begin{eqnarray}}   \nc{\eea}{\end{eqnarray}}
\nc{\baa}{\begin{array}}      \nc{\eaa}{\end{array}}
\nc{\bit}{\begin{itemize}}    \nc{\eit}{\end{itemize}}
\nc{\ben}{\begin{enumerate}}  \nc{\een}{\end{enumerate}}
\nc{\bce}{\begin{center}}     \nc{\ece}{\end{center}}

\def\sq2{\sqrt{2}}

\def\ph{\varphi}

\def\m4{m^4(\ph)}
\def\mn2{m_n^2}

\def\v5{V^{(5)}}

\def\baa{\begin{array}}
\def\eaa{\end{array}}

\begin{document}

\begin{flushright}
 \mbox{\normalsize \rm CUMQ/HEP 181}\\
\end{flushright}

\vskip 20pt

\title{Higgs Production in General 5D Warped Models}

\author{Mariana Frank\footnote{mariana.frank@concordia.ca}}
\affiliation{Department of Physics, Concordia University\\
7141 Sherbrooke St. West, Montreal, Quebec,\\ CANADA H4B 1R6
}
\author{
Nima Pourtolami\footnote{n\_pour@live.concordia.ca}}
\affiliation{Department of Physics, Concordia University\\
7141 Sherbrooke St. West, Montreal, Quebec,\\ CANADA H4B 1R6
}
\author{Manuel Toharia\footnote{mtoharia@physics.concordia.ca}}
\affiliation{Department of Physics, Concordia University\\
7141 Sherbrooke St. West, Montreal, Quebec,\\ CANADA H4B 1R6
}

\date{\today}

\begin{abstract}

We calculate the production rate of the Higgs boson at the LHC in the
context of general 5 dimensional (5D) warped scenarios with spacetime  background
modified from the usual $AdS_5$, and   where all the SM fields,
including the Higgs,  propagate in the bulk. This modification can
alleviate considerably the bounds coming from  precision electroweak
tests and flavor physics. We  evaluate the  Higgs  production rate and
show that it is generically consistent with the current
experimental results from the LHC for Kaluza-Klein (KK) masses as low
as 2 TeV, unlike in pure $AdS_5$ scenarios, where for the same masses,
the Higgs production typically receives corrections too large to be consistent
with LHC data. Thus the new pressure on warped models arising from LHC
Higgs data is also alleviated in $AdS_5$-modified warped scenarios. 

\end{abstract}


\maketitle

\section{Introduction}
\label{sec:intro}

While the recent discovery of a Standard Model (SM)-like Higgs boson at the LHC
completes the particle spectrum of the Standard Model (SM), from the
theoretical standpoint, the SM still seems incomplete.  Among other
things, no explanation is offered for the  hierarchy puzzles, one of
which concerns the large mass gap between the electroweak scale
($M_{\rm ew} \sim 200 $ GeV) and the Planck scale ($M_{\rm Pl}\sim 10
^{18}$ GeV). Another hierarchy is the one in the observed masses of
the fermions, from the very light neutrinos ($m_\nu \sim 5\times 10^{-2}$ eV)
to the top quark ($m_t \sim$ 175 GeV).  A popular modification to the
SM that tries to address these issues is to modify the space-time
symmetries by extending the number of space dimensions. If the
additional dimension, extending between two branes, one at the TeV
scale (IR brane) and the other at Planck scale (UV brane), with
gravity propagating in the bulk, is warped, the resulting geometry
generates naturally the Planck-electroweak scale hierarchy, with a
large $M_{\rm Pl}$ generated from a small length of the extra
dimension \cite{RS}. While in the original  Randall-Sundrum (RS) model
the SM fields were located on the IR brane, it was later shown that if
one lets the SM fields - except for the Higgs - to propagate in the
bulk of the  fifth dimension, fermion masses can be naturally
hierarchical, with masses determined by their localization with
respect to the two branes: the lighter fermions are localized near the
Planck brane, while the heavier ones are localized near the TeV
brane. The mass is determined by  the overlap integrals with the TeV
localized Higgs profile. This new framework was able to address issues
of the original RS model associated with flavor-changing neutral
currents (FCNC), proton decay and neutrino masses
\cite{Davoudiasl:1999tf, Pomarol:1999ad,  Grossman:1999ra,
  Chang:1999nh, Gherghetta:2000qt,   Davoudiasl:2000wi}.     

However, generic models with warped extra dimensions are still 
very constrained by electroweak and flavor precision tests 
\cite{Carena:2004zn, Huber:2003tu, Agashe:2006at, Csaki:2008qq} so
that the scale of the lightest KK  modes should be set to $\cal O$(10
TeV) or more. 
Various methods can be implemented to avoid some of these tensions.
To reduce pressure from electroweak precision tests, one can enlarge 
the gauge symmetry of the SM by introducing a custodial 
symmetry that limits the corrections to various precision
observables\cite{Agashe:2003zs,Agashe:2006at}.  Even with custodial
protection, very strong flavor constraints (specifically coming from
$K^0- \bar K^0$ mixing) must still be addressed \cite{Csaki:2008zd};
in the absence of any flavor symmetry\footnote{See for example
  \cite{Santiago:2008vq,Csaki:2008eh} for minimal flavor proposals
  managing to lift importantly the flavor bounds.} it was noted that
when the Higgs is allowed to leak out of  the TeV brane and its 5D
Yukawa couplings enhanced, there is a general reduction of flavor
bounds, but still keeping the KK masses at some $3-5$ TeV or more
\cite{Agashe:2008uz}.

Another interesting alternative to address tensions from  precision
electroweak and flavor tests is to modify the space-time metric, so
that close to the TeV brane, the background deviates from pure
five-dimensional anti-de Sitter space ($AdS_5$). This modification
suppresses large corrections to the electroweak and flavor observables
and makes it possible to reduce the constraints to $M_{KK}\gsim 1$
TeV, without the need to invoke custodial symmetry
\cite{Cabrer, Carmona:2011ib,MertAybat:2009mk}. A comprehensive analysis of the
implications of these models at the LHC, analyzing the production of
both electroweak and strong KK gauge bosons,  has been performed in
\cite{deBlas:2012qf}.   

However, it has also been pointed out that a new source of potential tension 
in $AdS_5$ scenarios can arise from the Higgs sector itself \cite{Neubert, Azatov:2010pf}.
The towers of fermion KK modes can affect significantly
the Higgs boson production rate by either enhancing or suppressing the Standard 
Model prediction. The predicted suppression or enhancement depends on
the model parameters considered, such as the nature of the Higgs (bulk
or brane localized), or the phases of the Yukawa operators and other
higher dimensional operators \cite{Azatov:2010pf,Frank:2013un}. It is interesting to note that in  
the case of a bulk Higgs the importance of higher dimensional
operators is reduced \cite{kaprivatecomm}, as well as the effect of the
phases in brane localized Yukawa operators. In that situation, one
obtains a more specific prediction for the effects on the Higgs
production rate, namely that there should be a general enhancement
with respect to the SM prediction \cite{Azatov:2010pf, Frank:2013un}.  
In this same scenario of bulk Higgs, the physical Yukawa couplings
between Higgs and fermions are overall suppressed \cite{Azatov:2009na,
  Frank:2013un}, both effects (enhancement in Higgs production and
suppression in fermion Yukawa couplings), being intimately
correlated\footnote{See \cite{Agashe:2009di} for a description of the
  same effect in the Yukawa couplings from the CFT dual picture.}.

Motivated by these considerations, we investigate what is the
situation for Higgs production in more general warped models with a
modified $AdS_5$ metric. Since some of these models manage to avoid all
precision tests for quite low KK scales (2 TeV) we ask the question of
whether, for such low KK masses,  they can also limit the potential
enhancements in Higgs production,  present in RS scenarios.  

Our work is organized as follows. In the next section,
Sec. \ref{sec:1}, we present a brief description of standard RS
scenarios and of two more general warped space scenarios. In Sec. \ref{sec:2} we give
the results for Higgs production through gluon fusion in the two models
with generalized warped space metrics and compare them with the RS
predictions. We then discuss the decoupling of the heavier modes in
Sec. \ref{sec:3} and finally we summarize our findings and conclude in
Sec. \ref{sec:summary}.  Some explicit formula are left for the
Appendix (Sec. \ref{sec:app}).

\section{Soft-wall inspired models}
\label{sec:1}
The setup of warped extra-dimensional models consists of a slice of a
five-dimensional anti-de Sitter space $AdS_5$, where the effective
$4D$ scale is dependent of the position of the extra dimension.  In
the standard RS formulation, the metric is given by 
\beq
ds^2=e^{-2 A(y)}\eta_{\mu \nu}dx^\mu dx^\nu-dy^2,   \  {\rm with}~A(y) =  ky, 
\eeq
where the two branes are localized at $y=0$ and $y=y_1$, $k \sim
M_{Pl}$ is the curvature scale of the $AdS_5$, and we are using the
mostly positive metric for  $\eta_{\mu \nu}$. Solving the gauge
hierarchy problem requires that the warping exponent, $k y_1$, to be
around $\sim 35$, which can be stabilized with a modest fined-tuning
of the parameters \cite{Goldberger:1999uk}. The TeV
scale is generated from ${\bar M}_{Pl}e^{-k y_1}$, with ${\bar
  M}_{Pl}$ the reduced $4D$ Planck scale.   
The KK excitations of the gauge bosons contributions to the
electroweak precision observables - especially the $T$ parameter -
introduce a lower bound on the masses of these KK excitations 
of about $\sim 10$ TeV, making them unobservable at LHC.

It was then observed that one way to address this issue was to
consider a stabilized solution to the 5D scalar-gravity system, in
which the $AdS_5$  behaviour near the UV brane was maintained,
but a deformation of conformality near the IR brane was apparent
\cite{Cabrer, Carmona:2011ib}. 
These scenarios assume a bulk Higgs and allow for a softening of
electroweak constraints through suppressed couplings of the electroweak
KK modes.  In turn, this  lowers the bounds on
the KK masses, yielding a model within the reach of the LHC in the near
future.  Assuming the following superpotential with real arbitrary
parameters $\nu$ and $b$  $$W = 6 k (1+ b e^{\nu \phi / \sqrt{6}}),$$
the stabilizing scalar field, $\phi$, and the modified metric warp
factor, $A(y)$ can be found,  and are given by \cite{Gubser:2000nd} 
\bea
\phi(y)=-{\sqrt{6}\over \nu}log(\nu^2bk(y_s-y))
\eea
\beq
A(y)=ky+\frac{1}{\nu^2} \ln \left ( 1- \frac {y}{y_s} \right), 
\eeq
where $y_s = y_1+\Delta$ ( $\Delta > 0$ ) is the position of the
singularity imposed by the scalar field, and is  {\it outside} of the
physical dimension,  so the logarithm is always single valued and
positive within the physical distance $y = 0$ and $y = y_1$. The RS
limit is obtained in either of the limits  $\nu \rightarrow \infty$,
or $y_s\rightarrow \infty$. The curvature radius along the extra
dimension is given by  
\bea\label{kL} 
L(y) = {\nu^2 (y_s - y)\over \sqrt{1-2\nu^2/5+2\nu^2k(y_s-y)+\nu^4k^2(y_s-y)^2}}.
\eea
The requirement that the gravitational expansion remains perturbative,
yields the following bound on this radius 
\bea\label{kLbound}
kL_1\equiv k L(y_1)\geq 0.2.
\eea
The bulk 5D Higgs profile along the extra dimension can be given by
\beq\label{higgs}
h(y)=h_0(y)e^{aky}, \qquad h_0(y)=\alpha_1 \left (1+ \alpha_2 \int_0^ye^{4A(y')-2aky'}dy' \right),
\eeq
where $a$ is a parameter that determines the localization of the Higgs profile along the extra dimension, which holographically can be viewed as the dimension of the Higgs condensate
operator. 
In order to solve the hierarchy problem (Higgs localized near the IR brane) we must have $a\geq a_{min}$. 
Following \cite{Cabrer} we introduce the following measure 
$$\delta\equiv \left|e^{-2(a-2)k y_s}ky_s(-2(a-2)k y_s)^{{4\over\nu^2}-1}\Gamma(1-{4\over\nu^2},-2(a-2)k(y_s-y_1))\right|$$
which is an estimate of the amount of fine tuning needed in order to save the RS solution to survive the hierarchy problem. To determine $a_{min}$
throughout this paper we have set $\delta=0.1$\footnote{We obtain, approximately, $\displaystyle a_{min} \simeq 2\frac{A(y_1)}{ky_1}$, and for a physically acceptable model we usually require 
$\displaystyle a \geq 2\frac{A(y_1)}{ky_1}$. The $\delta$ criterion given above usually corresponds to
an $a_{min}$ smaller than this estimation which in turn leads to some
fine tuning of the 5D parameters.} in the above equation and solved
for $a$.  The profile $h_0$ in equation (\ref{higgs}) is given by the
normalization condition $$\int_0^{y_1}dyh(y)^2e^{-2A(y)}=1,$$ and
obtained explicitly as 
\begin{eqnarray}\nonumber
h_0&=&\left\{e^{2(a-1)ky_s}(2(a-1)ky_s)^{-(1+{2\over\nu^2})}\right.\\
&&\left.\left(\Gamma\left[(1+{2\over\nu^2}),2(a-1)k(y_s-y_1)\right]-\Gamma\left[(1+{2\over\nu^2}),2(a-1)ky_s)\right]\right)y_s\right\}^{-{1\over2}}.   
\end{eqnarray}
The matter action of the SM fields in Dirac spinor notation is given by
\begin{eqnarray}
S_{mat} = \int d^5x\sqrt{-g} {\cal L}_{mat} =  \int d^5x\sqrt{-g}
({1\over2}(i\bar{\Psi}\Gamma^M D_M \Psi - iD_M\bar{\Psi}\Gamma^M\Psi)
+ M_\Psi(y){\bar\Psi}\Psi), 
\end{eqnarray}
where $\Psi=(\psi_L,\psi_R)^T$ are the 5D fields and the capital
index, $M$, runs over the five space time dimensions with the
spinor fiber indices being summed over as $\Gamma^M\equiv
E^M_a\gamma^a$, with $\gamma^a\equiv(\gamma^\mu,\gamma^5)$. The
covariant derivative is $D_M=\partial_M+\omega_M$ with the spin
connection given by
$\omega_M={1\over8}\omega_{MAB}[\gamma^A,\gamma^B]$. The funfbein and
the inverse funfbein are given by  
$$e^a_M=(e^{-A(y)}\delta_\mu^\alpha, 1),\ E^M_a=(e^{A(y)}\delta^\mu_\alpha, 1).$$ 
The mass term coefficient, $M_\Psi(y)$ in general depends on the extra
dimension coordinate and  $\sqrt{-g}=e^{-4A(y)}$.  
Following the standard ans\"atz, we decompose the fields into an extra
dimensional profile field and SM $4D$ fields. 
The equations of motion of the profiles can be decoupled and written in the following convenient form
\bea\label{swdeu}
\partial_y(e^{-A-2Q}\partial_y(e^{Q-2A}\psi_L))+m_n^2e^{-Q-A}\psi_L=0,
\eea
\bea\label{swdeq}
\partial_y(e^{-A+2Q}\partial_y(e^{-Q-2A}\psi_R))+m_n^2e^{Q-A}\psi_R=0,
\eea
where we defined
\bea\label{bm1}
Q(y)\equiv\int_0^yM_\psi(y')dy'.
\eea
We are going to consider two different choices for the above fermion bulk mass, $M_\psi$:
 \begin{itemize}
\item Inspired by the standard RS choice $Q^{RS}_\psi(y) = c_\psi ky$ one can chose a
$y$-dependent bulk mass proportional to the warp factor $A(y)$. We
  call this the CGQ scenario  \cite{Cabrer} 
\bea\label{bm2}
Q_\psi(y)=c_\psi A(y)
\eea
\item Alternatively one can consider a constant mass, and we denote
  this choice as the CPS scenario   \cite{Carmona:2011ib}, (see also \cite{MertAybat:2009mk})
\bea\label{bm3}
Q_\psi(y)=c_\psi k.
\eea
\end{itemize}
As usual, imposing the proper boundary conditions on either left or
right handed fields will ensure the existence of normalized chiral
massless zero modes. Their expressions are: 
\bea\label{u0}
{\rm[CGQ]}\ \ && \ \left\{\begin{array}{ccc}
u_R^0&=&f(-c_u)\ e^{(2+c_u)A(y)} \\
q_L^0&=&f(c_q)\ e^{(2-c_q)A(y)}
\end{array}\right.\\
\label{q0}
{\rm[CPS]}\ \ && \ \left\{\begin{array}{ccc}
u_R^0&=&f(-c_u)\ e^{(2+c_u)ky}(1-{y\over y_s})^{-{2\over \nu^2}}\\
q_L^0&=&f(c_q)\ e^{(2-c_q)ky}(1-{y\over y_s})^{-{2\over \nu^2}}
\end{array}\right.
\eea
and we have defined\footnote{In analogy with the usual RS profiles  
$f(c)\equiv\sqrt{\frac{1-2 c}{1-e^{(1-2 c)k y_1}}}$.}
\begin{eqnarray}\nonumber
f(c)&\equiv&
\left\{y_s\big((1-2 c)ky_s\big)^{\frac{1-2 c}{\nu ^2}-1} e^{(1-2c) k y_s}
\vphantom{\int^\int}\ \times\right.\\
&&\left.\left[\Gamma\left(1-\frac{1-2 c}{\nu^2},(1-2 c) k(y_s-y_1)\right)-
\Gamma\left(1-\frac{1-2c}{\nu ^2},(1-2 c) ky_s\right)\right]\right\}^{-{1\over2}}
\end{eqnarray}
for the CGQ scenario. For the case of the CPS scenario one just needs
to set $c=0$ $inside$ the gamma functions, and leave everything else as is.

For generic values of $\nu$ and $y_s$ the eigenvalues and
eigenfunctions (masses and profiles) of the KK fermion modes can be obtained
by solving numerically equations (\ref{swdeu})  
and (\ref{swdeq}).

\section{Higgs Production through gluon fusion}
\label{sec:2}

In this section we solve numerically for the eigenvalues and
eigenfunctions of Eqs. (\ref{swdeu}) and (\ref{swdeq}) with the goal 
 of calculating the Higgs production rate through gluon fusion.
(Analytic solutions are unfortunately not available).

\begin{figure}[t]
\center
\begin{center}

	\includegraphics[height=4cm]{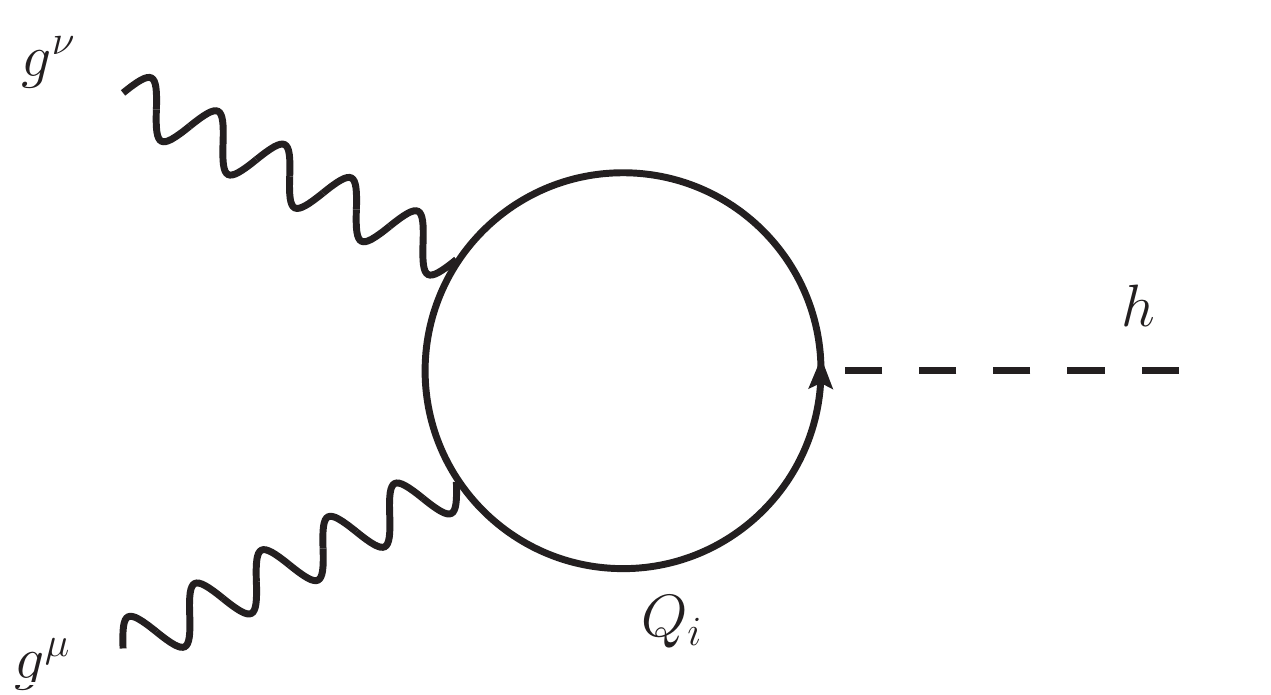}
 \end{center}
\caption{Loop diagram showing the contribution of the quark $Q_i$ to the
  Higgs-gluon-gluon coupling. In the SM, the dominant contribution is
  through the top quark due to its large Yukawa coupling with the
  Higgs boson. In warped space models the heavier KK fermions contribute to the coupling
  with potentially large effects, either suppressing or enhancing the
  SM coupling, depending on the phases present in the different
  Yukawa-type operators present in the 5D action, and on the
  localization of the Higgs (see text for details).} 
\label{fig:toploop}
\end{figure}

In the SM, the main contribution to the Higgs coupling to gluons comes
from a top quark loop correction. In warped extra dimensional models there are
many heavy KK quarks with important couplings with the Higgs, and thus one needs to
add all of their contributions to the loop (see Figure
\ref{fig:toploop}), so that the resulting cross section for the process $gg\to h$ is  
\bea
\sigma^{SM}_{gg\rightarrow h} = {\alpha_s m_h^2\over
  576\pi}\left|\sum_Q{y_Q\over m_Q}
A_{1/2}(\tau_Q)\right|^2\delta({\hat s}-m_h^2)\label{hggSM},  
\eea
with $\tau_Q\equiv m^2_h/4m_Q^2$, $\hat{s}$ being the $gg$ invariant
mass squared and $Q$ representing the physical fermions with physical Yukawa couplings
$Y_Q$ and masses $m_Q$. The form factor is given by  
\bea
A_{1/2}(\tau) = {3\over 2}[\tau +(\tau -1)f(\tau)]\tau^{-2}\label{ff},
\eea
with
\bea
f(\tau) = \begin{cases}
      [\text{arcsin}\sqrt{\tau}]^2  \qquad\;\qquad\;\qquad{\tau\leq 0}\\
      -{1\over 4}\left[\text{ln}\left({1+\sqrt{1-\tau^{-1}}\over
          1-\sqrt{1-\tau^{-1}}}\right)-i\pi\right]^2
      {\tau>1}\label{ftau}. 
          \end{cases}
\eea   
The relevant quantity that we wish to evaluate is the effective coupling
\bea\label{cgg1} 
c_{hgg}=\sum_Q{y_Q\over m_Q} A_{1/2}(\tau_Q),
\eea 
where $y_Q$ is the $4D$ Yukawa coupling of fermion Q, in the mass eigenbasis, and 
$m_Q$ is its mass. In the gauge basis (the basis {\it before} spontaneous
EW symmetry breaking)  the Yukawa couplings are given by the following
overlap integral along the fifth dimension, 
\bea
Y_{Q_iU_k}^u = Y^u\int_0^{y_1}dye^{-4A(y)} h(y) Q_L^{u,(i)}(y)U_R^{(k)}(y)\label{yu},
\eea
and can be written as the following infinite dimensional Yukawa matrix, $\mathbf{Y}$
\bea\label{Y}
\begin{pmatrix}
q_L^{0},\ Q_L^{i},\ U_L^{j}
 \end{pmatrix}
\begin{pmatrix}
   Y^u_{{q}_Lu_R}\         &          0            &   Y^u_{{q}_LU_R^b}\ \\ 
   Y^u_{{Q}_L^iu_R}\         &          0         &   Y^u_{{Q}_L^iU_R^b}\ \\ 
        0                  &  Y^{u*}_{{U}_L^jQ_R^a}\    &     0 
\end{pmatrix}
\begin{pmatrix}
u_R^{0}\\ Q_R^{a}\\ U_R^{b}
 \end{pmatrix}.
\eea 
In the same gauge basis we can also write down the infinite
dimensional fermion mass matrix 
\bea\label{M}
\mathbf{M}=
\begin{pmatrix}
   Y^u_{{q}_Lu_R}\  v_4       &          0            &   Y^u_{{q}_LU_R^b}\ v_4\\ 
   Y^u_{{Q}_L^iu_R}\  v_4       &          M_Q          &   Y^u_{{Q}_L^iU_R^b}\ v_4\\ 
        0                  &  Y^{u*}_{{U}_L^jQ_R^a}\ v_4   &      M_U 
\end{pmatrix}.
\eea 
where $v_4=174$ GeV is the Higgs vacuum expectation valued (VEV), and
$M_Q$ and $M_U$ are the $n$-dimensional diagonal matrices of the tower
of $n$ KK modes mass eigenvalues.

We proceed here by considering an effective field theory with only 3 KK
levels, but we refer the reader to Section IV for a discussion
involving the use of the full tower of KK fields.
In our effective approach here, the previous  matrices
$\mathbf{M}$ and $\mathbf{Y}$  become $7\times 7$ truncated mass and
Yukawa matrices. In order to use  Eq. (\ref{cgg1}), we diagonalize
numerically the mass matrix $\mathbf{M}$  above by a bi-unitary
transformation. Performing the same transformation on the Yukawa
matrix, $\mathbf Y$,  one can finally use  Eq. (\ref{cgg1}) to
calculate the Higgs production cross  section\footnote{If flavor
  families are included, this transformation in general will not diagonalize the full Yukawa
  matrix $\mathbf{Y}$ leading thus to tree-level Higgs    mediated flavor
  changing currents \cite{Agashe:2009di,Azatov:2009na}.}.   

In all our numerical analysis we took $k=10^{18}$ GeV fixed, and then
tuned all other parameters in such a way that all of the scenarios yield
approximately the same zero mode masses (the SM quark masses) and the same
lightest KK mode mass ($\simeq 2.1$ TeV). While this value for the
lightest KK mass is already dangerously low for RS scenarios, in
general warped models with modified $AdS_5$ metrics (like the CGQ and
CPS scenarios considered here), such low KK masses can be safe from electroweak precision
constraints as well as flavor constraints \cite{Cabrer, Carmona:2011ib}. The regime in which this happens is such that
$kL_1\simeq 0.2$, and so we restrict ourselves to that region of parameter space
when considering the modified models.

\begin{figure}[t]
\center
\begin{center}

	\includegraphics[height=9.5cm,width=5.6cm]{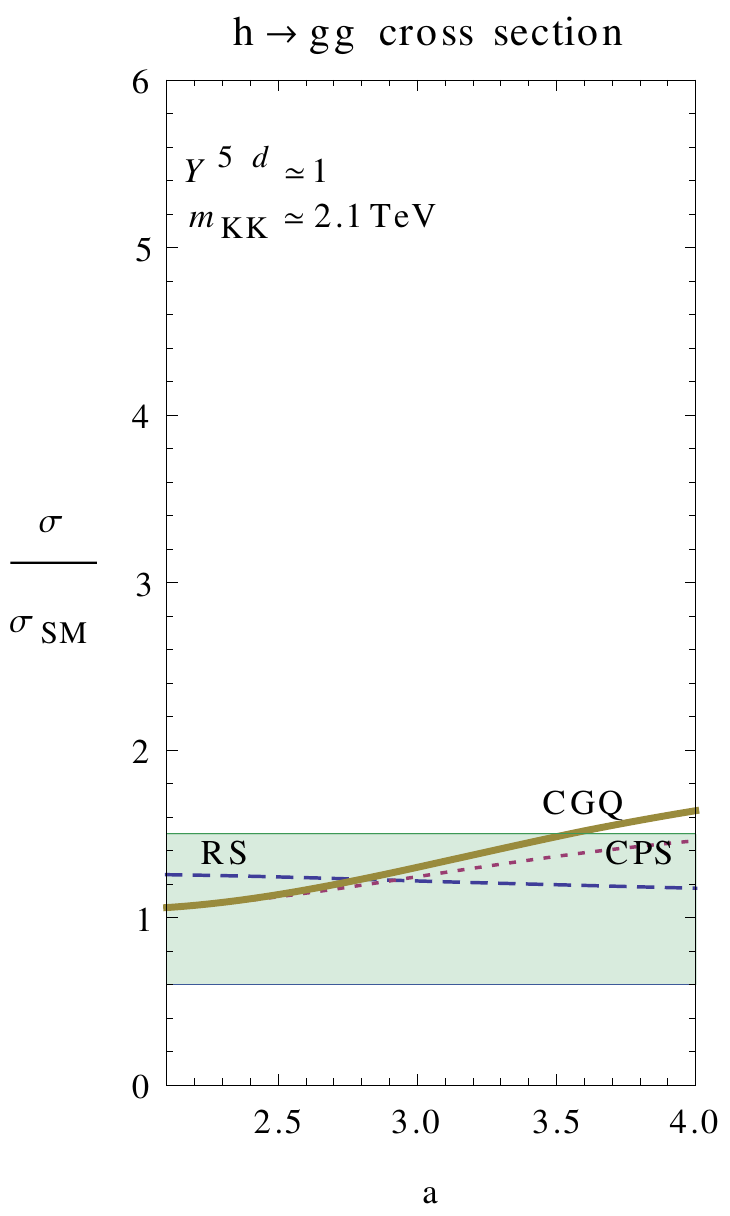}\hspace{-.3cm}
	\includegraphics[height=9.5cm,width=5.6cm]{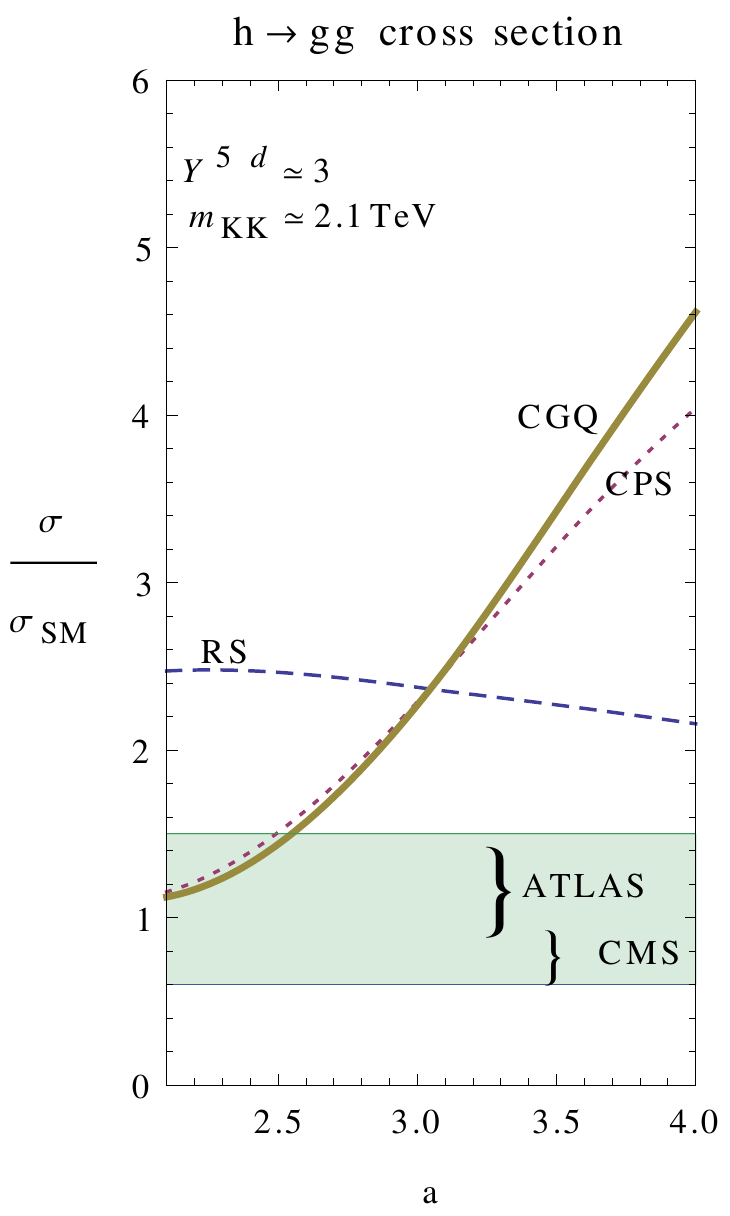}\hspace{-.3cm}
	\includegraphics[height=9.5cm,width=5.6cm]{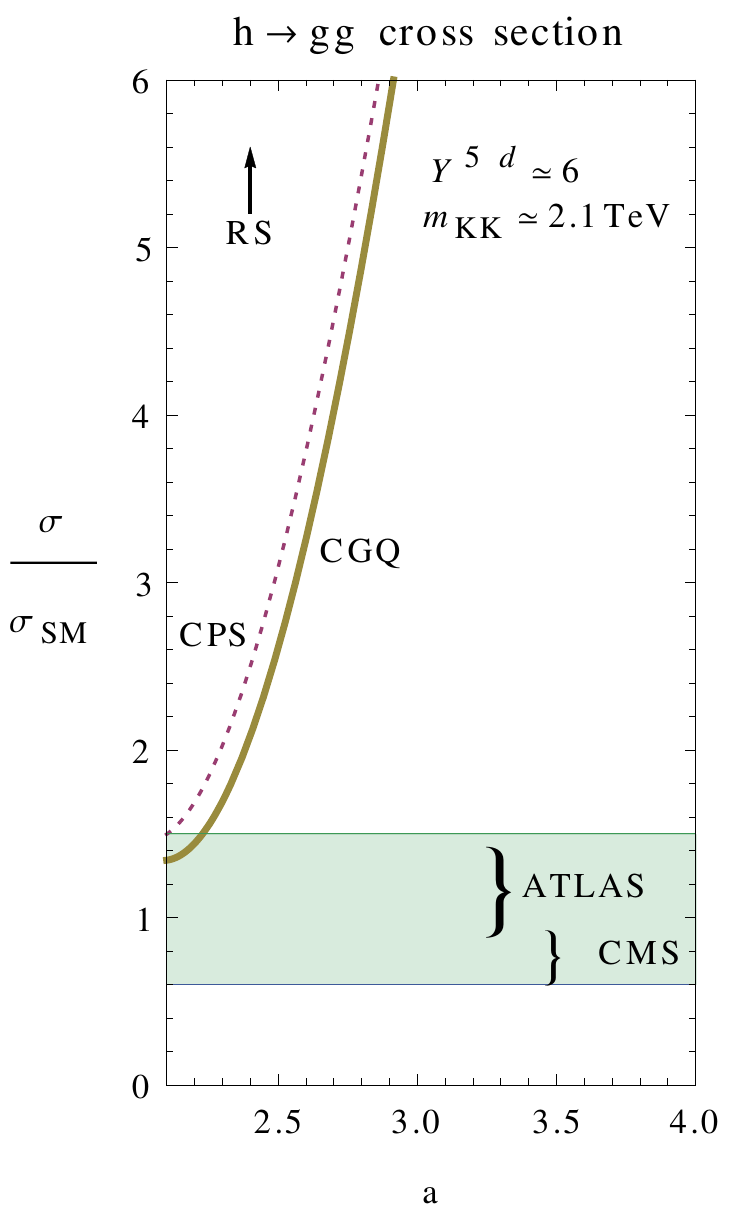}
 \end{center}
\caption{Higgs production rate ratio to Standard Model
  prediction as a function of the Higgs localization parameter,
  $a$. In all scenarios we consider an effective field theory consisting of a
  tower of 3 KK modes. The dashed line (blue) corresponds to the RS
  scenario with bulk Higgs. The solid and dotted lines correspond to
  the modified $Ads_5$ scenarios CGQ ($\nu= 0.5$, $kL_1\simeq 0.27$)
  and CPS ($\nu= 0.5$, $kL_1\simeq 0.29$) and the lightest KK mass is
  about $2.1$ TeV in all models. The 5D Yukawa coupling is varied
  between $Y^{5D}\sim1$ (left panel), $Y^{5D}\sim3$ (middle panel) and
  $Y^{5D}\sim 6$ (right panel). The shaded regions show the
  experimental bounds from CMS and ATLAS.
 }  \label{fighgg}
\end{figure}

It is important to note that the zero mode masses are sensitive to the
values of the Higgs localization parameter $a$ through  Yukawa couplings
\bea
y = {Y^{5D}\over\sqrt{k}}\int_0^{y_1}dye^{-4A(y)} h(y) q_L^{0}(y)u_R^{0}(y)\label{yu0}.
\eea
Using equations (\ref{higgs}), (\ref{u0}) and (\ref{q0}) the above integral can be evaluated
\begin{eqnarray}\nonumber
y&=&-\frac{h_0}{f(c_q)f(c_u)}y_s
   ((a-c_q+c_u)(ky_s))^{\frac{c_u-c_q}{\nu ^2}-1}
   e^{k y_s (a-c_q+c_u)}\times \\\nonumber
&&   \left(\Gamma \left[(\frac{c_q-c_u}{\nu^2}+1),(a-c_q+c_u) k y_s\right]-\Gamma
   \left[(\frac{c_q-c_u}{\nu^2}+1),(a-c_q+c_u) k(y_s-y_1)\right]\right).
\end{eqnarray}
In order to keep the 5D Yukawa couplings $Y^{5D}$ fixed for all of
these models\footnote{For a fair comparison among scenarios, we
   maintain the value of $Y^{5D}$ fixed in all of them, given
  that the production cross section is proportional to $(Y^{5D})^2$.}  while
requiring the correct SM quark masses, we have to set the
values of $c_q$ and $c_u$ separately for each value of $a$, and for
each different scenario.
On a technical level, to be able to produce a correct top quark mass we had to set
$Y^{5D}_{top}\simeq 3$ in all cases (so that even when we write $Y^{5D}\simeq 1$  in the left
panel of Figure \ref{fighgg} we are still using $Y^{5D}_{top} \simeq 3$.).

To proceed further we need to include the flavor families of the SM,
and to do so we will consider a simplified version of the SM model,
that is, we take a $SM-like$ setup in which the 5D Yukawa couplings are
diagonal, and thus we will ignore flavor inter-mixings in our loop calculation. 
With this simplifying assumption, it is straightforward to add  
the contributions of all the fermions running in the loop. This is of course 
 not a viable scenario  of flavor, but it does illustrate
fairly the effects on Higgs production.  Thus we consider the presence
in the loop of five light quarks (which have negligible effect) and their
associated KK quarks (which yield the main new contributions), as
well as one SM top quark and its associated 
KK top quarks (the overall contribution from the top sector is
actually very close to the SM top contribution). This simplified
flavor structure for fermions is used in the three scenarios we consider, 
i.e. bulk-Higgs-RS, CGQ and CPS, and so we can obtain fair comparisons
between the different predictions for Higgs production cross 
section generated by each model.

In the case of the CGQ scenario, the fermion bulk mass term is $M_i=c_i A'(y)$
(see Eqs. (\ref{bm1}) and (\ref{bm2})) and we used the values $\nu= 0.5$, and
$kL_1\equiv kL(y_1)=0.27$ for the curvature radius  as defined in
equation (\ref{kL}). The $\delta= 0.1$ criterion yields $a_{min}\simeq
2.09$ and the values for $c_q$ ($\simeq-c_u$) are being slightly
decreased as $a$ becomes larger  in order to keep the $Y^{5D}$ constant (as explained above). 
The CPS case is essentially the same model but with a constant fermionic bulk
mass of $M_i=c_i k$ (see Eqs. (\ref{bm1}) and {\ref{bm3})). We have again used the value
$\nu=0.5$, but in order to keep the lightest KK mass fixed at the
same value as the other models we set $kL_1\equiv kL(y_1)= 0.29$ for
the curvature radius.  The $\delta=0.1$ criterion this time yields
$a_{min}\simeq 2.06$.  
Finally the RS scenario (with a bulk Higgs) is obtained in the limit
$\nu>>1$ and $kL_1\rightarrow 1$. The numerical results obtained in this
limit match the results obtained using the analytical RS formulae \cite{Azatov:2010pf,Frank:2013un},
providing thus a nontrivial consistency check of the procedure.

The results are shown in Figure \ref{fighgg}, in which we plot the
ratio of Higgs production cross section relative to the SM one, as a function of  the Higgs
localization parameter $a$ for the three models.  The solid and dotted
lines correspond to predictions from the CGQ and CPS models, respectively,
while the dashed line represents the prediction of the RS scenario
(with fermions and Higgs in the bulk). For each model, the numerical
calculation involves the contributions from light quarks and 3 KK
modes in the loop. The three panels correspond to
different values for the 5D Yukawa coupling: $Y^{5D}\sim 1$ on the
left side, $Y^{5D} \sim 3$ in the middle, and 
$Y^{5D}\sim 6$ on the right side. 
The shaded regions represent experimental restrictions on cross
sections from CMS  ($\mu\equiv{\sigma\over\sigma_{SM}}= 0.8\pm 0.22$) and ATLAS
  ($\mu\equiv{\sigma\over\sigma_{SM}}=1.2\pm 0.3$) results  \cite{Beringer:1900zz}.

From all panels, it is clear that the behaviour of the CGQ and CPS
models is very similar. In particular, both are very sensitive to the
Higgs localization parameter $a$, while the RS model is much more
stable against variations in $a$. Second, one can see that both
modified warped models alleviate the enhancements in Higgs
production present in RS models, but only for a small region of $a$,
fortuitously the same region for which the electroweak constraints are
satisfied \cite{Cabrer}. The restriction becomes more stringent
with increased $Y^{5D}$, so that for $Y^{5D} \sim 1$, the $a$ parameter can
be anywhere between its minimal value $a_{min}$ and about $3-4$ (depending on the
CGQ model or the CPS model), while for $Y^{5D}\sim 6$ the parameter $a$ is constrained to be in a
really small region around $a_{min}$.
By comparison, the RS model seems ``safe'' when $Y^{5D} \sim 1$\footnote{Of course, for such
  low KK masses the minimal RS scenario without custodial protection
  is already excluded due to precision electroweak tests and flavor
  bounds. In fact the smaller the value of $Y^{5D}$ the worse the flavor
  bounds become \cite{Csaki:2008eh,Agashe:2008uz}.
},  but then is completely disfavoured for $Y^{5D}\sim3$ and $Y^{5D}
\sim 6$ (where it lies outside the range of the figure).

The main message from these plots is that in the modified
warped scenarios, the Higgs should be as de-localized as
possible otherwise, if the Higgs is pushed towards the IR, the
bounds become even worse than in RS (or at least the bounds we have
considered here, namely those coming from LHC Higgs production).
The reason for this behaviour with the parameter $a$ is that in the CGQ
and CPS scenarios, the warp factor grows faster than in RS near the IR
boundary. The effect of this is to actually concentrate the KK modes
closer to the IR brane, and if one de-localizes sufficiently the Higgs
profile away from the IR, the overlap integral between Higgs and KK
fermions is suppressed with respect to the RS case. This leads to
suppressed corrections to Higgs production, and we believe that
this is also the origin for the suppressed contributions to
electroweak and flavor observables.

\begin{figure}[t]
\center
\begin{center}
	\includegraphics[height=9.5cm,width=8.3cm]{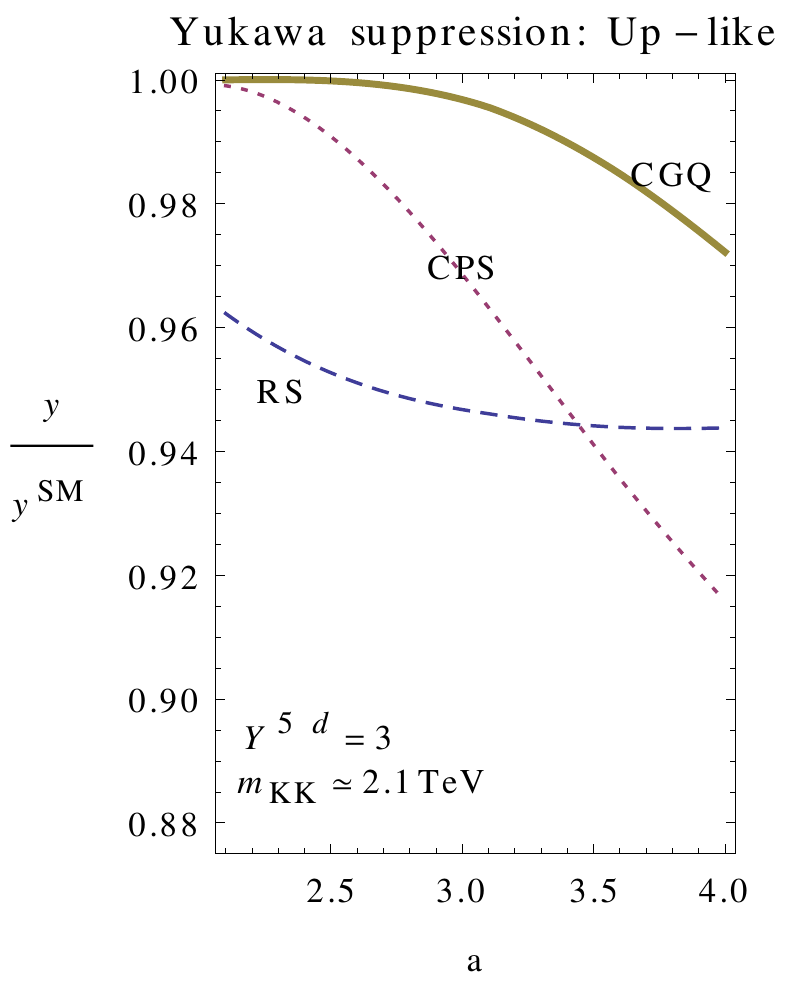}\hspace{-.7cm}
	\includegraphics[height=9.5cm,width=8.3cm]{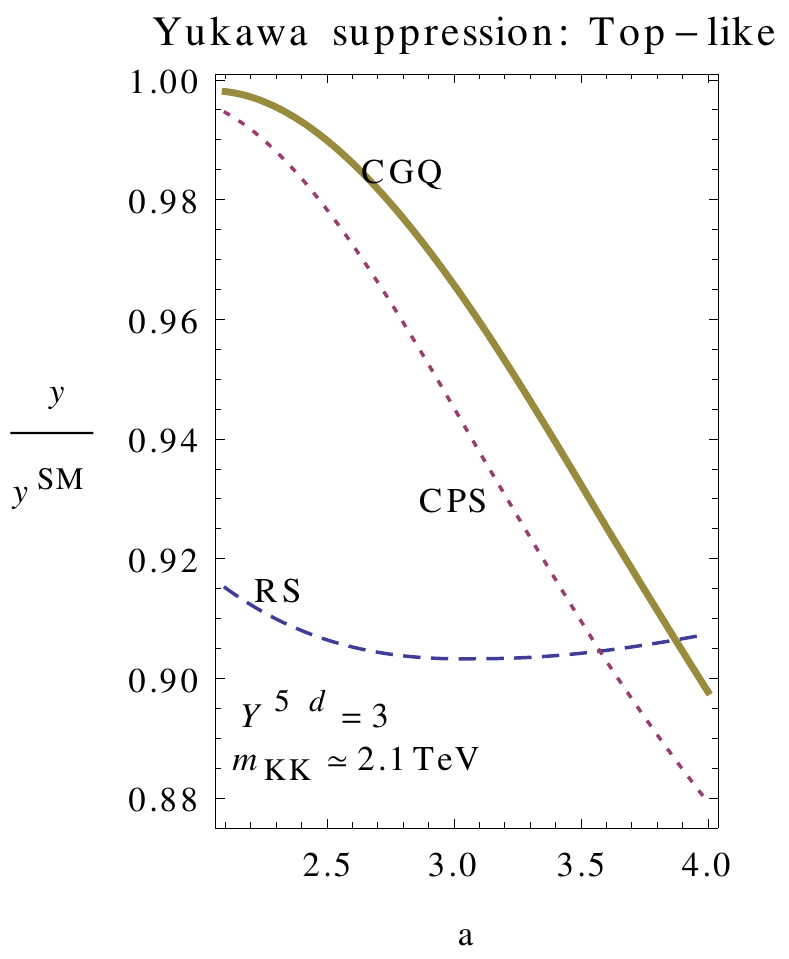}
 \end{center}
\caption{Quark Yukawa couplings relative to their SM values, of a light quark
  (left panel) and the top  quark (right panel) as a function of the Higgs
  localization parameter, $a$.  In all scenarios we consider an
  effective field theory consisting of a tower of 3 KK 
levels with $Y^{5D}\sim 3$ and with the lightest KK mass at about $2.1$  TeV. 
 The dashed line (blue) corresponds to RS with bulk Higgs. The
  solid and dotted lines correspond to the CGQ ($\nu= 0.5$,
  $kL_1\simeq 0.27$) and CPS ($\nu= 0.5$, $kL_1\simeq 0.29$)
  models of general metrics respectively.      } 
\label{figy}
\end{figure}

As a further check, in Figure \ref{figy} we plot
the relative size of light quarks Yukawa couplings (left panel)
and top quark Yukawa coupling (right panel) as a function of the
Higgs localization parameter $a$, where both effects are computed by
considering a truncated tower of just 3 KK fermion modes for each
scenario. We choose $Y^{5D} \sim 3$ in all 
plots and here again we observe that the suppressed Yukawa couplings in the RS
model are relatively independent of the Higgs de-localization, while in
both  the CPS and the CGQ scenarios, the suppression in these
couplings depends dramatically on the localization parameter $a$. 
These effects confirm the findings of the previous figure, namely that
for small $a$ (Higgs very de-localized) the modified warped scenarios
produce very little effects in the Higgs sector, but these effects
grow very quickly as the Higgs is pushed towards the TeV brane. The RS
dependence on Higgs localization is much milder, but the effects are
quantitatively quite large for the low KK masses considered here.
We  note here that even though the top Yukawa coupling can be quite
suppressed, the scenarios still predict an enhancement in Higgs
production. The reason is that the reduction produced by suppressed
top couplings is balanced by the positive contribution due to the top KK tower. To
this contribution, one must add the positive contributions of the other 5 towers of
KK fermions (associated to the 5 other SM quarks) \cite{Azatov:2010pf}.

Finally let us comment again that  we have focused on a
simplified version of the SM in which we ignored the flavor mixing between families, in
order to simplify the sums in the loop calculation. Our main goal was to compare Higgs production among
different models and study the general effects of the metric
modification relative to the usual RS setup. A realistic scenario
including the full flavor structure is underway, but the results
should not be much different from the ones presented here, since
flavor inter-mixings are not expected to produce big changes in the
contributions to the loops generating the  $hgg$ coupling.

\section{Decoupling of the Heavy KK modes}
\label{sec:3}
In this section, we consider the effect of including the full tower of
the KK modes on the evaluation of the Higgs production cross section,
by  performing the infinite sums analytically. Following the procedure
given in \cite{Frank:2013un,Azatov:2010pf} we obtain the following
expression for $c_{hgg}$, Eq. (\ref{cgg1}), 

\begin{eqnarray}\label{cgg2}
c_{hgg}&=&  
\text{Tr}(\mathbf{YM}^{-1}) + \sum_{\rm light}{y_Q\over
  m_Q}(A_{1/2}(\tau_Q) - 1)\label{sumym}\\
  &=&-2{v_4}\sum_{i,j}{Y^u_{{Q_L}_i{U_R}_j}Y^{u*}_{{U_L}_j{Q_R}_i}\over
  M_{Q_i}M_{U_j}} + {y_Q\over  m_Q}\big|_{\text{light}}A_{1/2}(\tau_{Q_{\rm light}}).
\end{eqnarray}
Here the couplings are the elements of the Yukawa matrix $\mathbf{Y}$ given in Eq. (\ref{Y}) and
$\mathbf{M}$ is the fermion mass matrix in the gauge basis given by Eq. (\ref{M}). We have also used the fact
that $\mathbf{Y}={\partial \mathbf{M}\over\partial{v_4}}$ and therefore
$\text{Tr}(\mathbf{YM}^{-1})=\text{Tr}({\partial \mathbf{M}\over\partial{v_4}}\mathbf{M}^{-1})={\partial \text{ln Det}(\mathbf{M})\over \partial v_4}$.

Since the form factor, $A_{1/2}$ is negligible for the light fermion
generations, we can neglect the last term in Eq. (\ref{cgg2}), and
using Eq. (\ref{yu}) we have
\bea\label{hggyu}
c_{hgg}=-2{v_4}Y^uY^{u*}\sum_{i,j}\int dydy'e^{-4A(y)}e^{-4A(y')}
{Q^{(i)}_L(y)Q^{(i)}_R(y')\over
  M_{Q_i}}{U^{(j)}_R(y)U^{(j)}_L(y')\over M_{U_j}} h(y)h(y'),\ \ \ \ 
\label{C}
\eea
where the Higgs profile is given in Eq. (\ref{higgs}). 
The infinite sums in this equation can be performed using the completeness of the Sturm-Liouville system.
We have (see the Appendix)
\bea\label{sumu}
\sum_{n=1}^{\infty}{{\hat U}_R^{(n)}(y){\hat U}_L^{(n)}(y')\over m_n}=e^{Q(y)-Q(y')}\left[\theta(y'-y)-{\int_0^{y'}e^{A-2Q}\over\int_0^{y_1}e^{A-2Q} }\right]
\eea
\bea\label{sumq}
\sum_{n=1}^{\infty}{{\hat Q}_L^{(n)}(y){\hat Q}_R^{(n)}(y')\over m_n}=-e^{Q(y')-Q(y)}\left[\theta(y'-y)-{\int_0^{y'}e^{A+2Q}\over\int_0^{y_1}e^{A+2Q} }\right].
\eea
Inserting this back into  Eq. (\ref{hggyu}) for $c_{hgg}$ we can finally calculate the $hgg$ cross section.
\begin{figure}[t]
\center
\begin{center}
	\includegraphics[height=8cm]{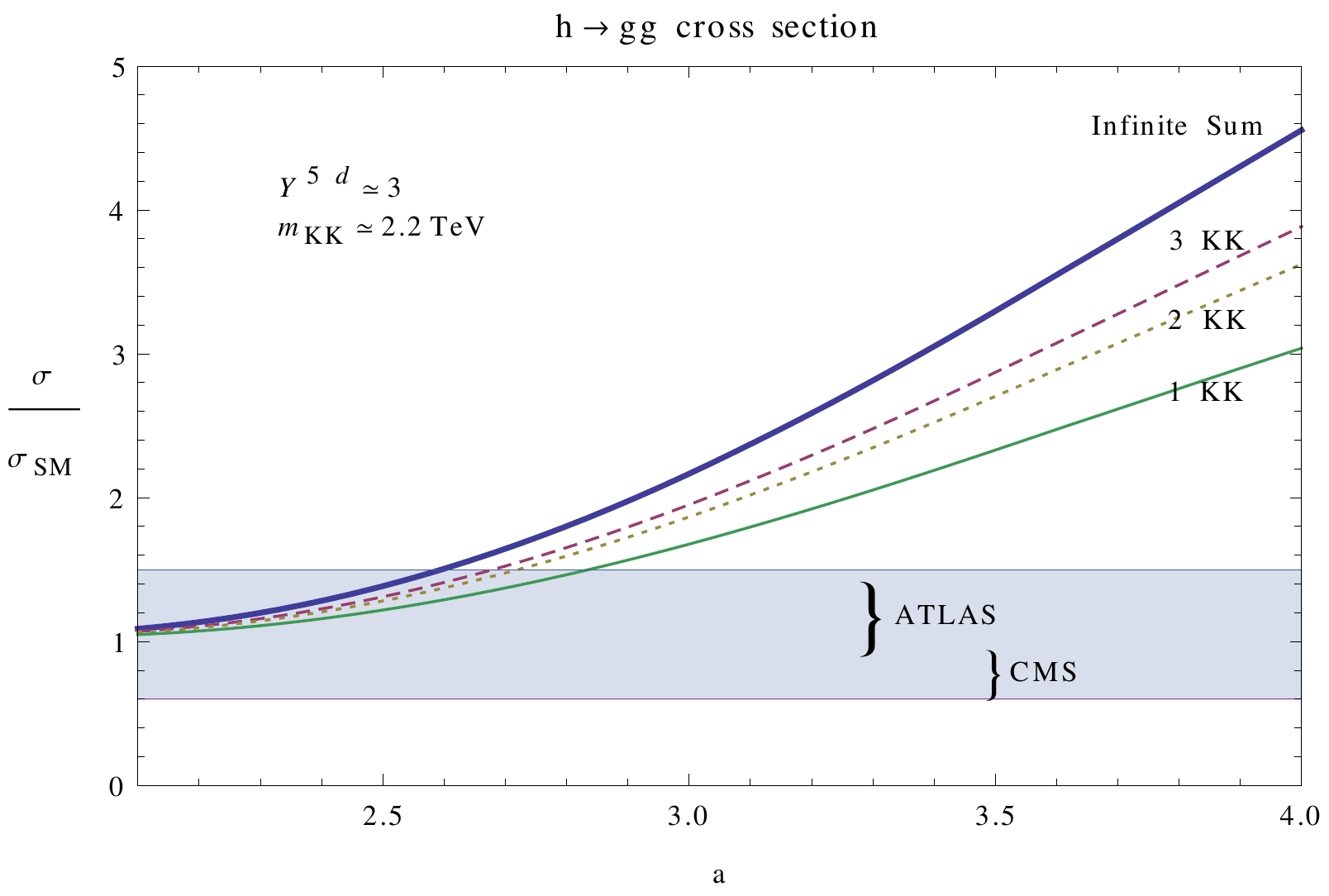}
 \end{center}
\caption{Higgs production rate via gluon fusion in the CGQ
  scenario. The solid thick line (blue) shows the contribution due to
  the full tower of KK modes. The dashed (red), dotted (khaki) and
  thin solid (green) lines show the contribution due to a tower of 3,
  2 and 1 KK modes respectively. The shaded region shows the
  experimental bounds from CMS and ATLAS.} 
\label{figinf}
\end{figure}
Figure \ref{figinf} shows the result of evaluating the Higgs
production cross section using the infinite KK tower, compared to the
result obtained using effective field theories with one, two and three
KK modes. As we can see, the result obtained  using a very small
number of modes converge quickly to the result using infinite sum,
which means that the heavy KK modes decouple from the evaluation of
the cross section. The decoupling of heavy modes is particularly true
in the region where the Higgs production cross section is safe from
large enhancements from the presence of KK modes in the loop. This
observation seems to imply that at least for the observable considered
here, the Higgs production rate, the 5D warped models with bulk
Higgs, which are essentially valid only up to $\Lambda_{UV}\sim
\cal{O}(\text{10 TeV})$, are calculable and the effects
of higher order operators should be suppressed (see also \cite{Frank:2013un}).

\section{Conclusions}
\label{sec:summary}
In this work we calculated the Higgs production rate via gluon fusion
in 5D scenarios with modified $AdS_5$ metrics.
In the SM, the Higgs production cross section is determined by the coupling of the
top quark to the Higgs field. Just like in RS, the deviations from the SM in modified $AdS_5$ 
models are caused  by the presence of extra KK fermion towers, associated to
each of the six SM quarks, and their couplings to the Higgs field. These KK fermions circulate
through  the loop responsible for the Higgs production through gluon
fusion, and they can lead to large enhancements in the Higgs production
cross section.
In RS models with fermions and Higgs fields propagating in the bulk, depending on the values of the
5D Yukawa couplings, the enhancements can reach 50\%
if the lightest KK mass is $\sim 2$ GeV (at odds with ATLAS
and CMS data). In fact the new data from LHC Higgs searches have become a
stringent bound on warped scenarios, to be considered together with
flavor and electroweak precision tests.

The couplings with the KK fields are generated by the
overlap of the KK fermions wavefunctions and the Higgs profiles along the fifth dimension, thus
the localization of these fields is crucial to the calculation. 
In the general warped scenarios, the KK fermions are pushed {\it towards} the
IR brane compared to the RS model due to the metric growth near that boundary. On the other
hand the localization of the Higgs profile along the 5th dimension is
controlled by a free parameter $a$: the smaller the values for $a$,
the less localized the Higgs profile\footnote{As we mentioned before,
  holographically this parameter corresponds to the dimension of 
the Higgs condensate operator and therefore its value is crucial in obtaining the 
correlation functions in the strongly coupled {\it modified}-CFT theory.}.
Our results show that a more de-localized Higgs field leads to a more
SM-like Higgs production, due to suppressed overlap integrals between
Higgs and KK fermions. Moreover, this seems to happen in the parameter
region which is also safe from electroweak and flavor precision
tests. We have also shown, by comparing the results obtained using a
small number  of modes with  the effects of the whole KK tower, that
the former converges quickly to the latter in the same region,
leading further support to the validity of our calculation.

Thus we have shown that, based on results from Higgs production, the
modified $AdS_5$ scenarios considered here are consistent
with the experimental results for light KK masses of $\sim 2$ TeV (unlike RS models). 
This was a non-trivial check, and necessary, since while these scenarios had
proved to be safer in terms  of precision tests (electroweak and
flavor) compared to RS,  the new Higgs production data from the LHC might have
been in conflict with the effects from the models. Our results confirm the viability of these
scenarios, which allow for new physics at lower scales than
conventional RS models and thus could yield signals at the LHC in the
near future.

\section{Acknowledgements}
We thank NSERC for partial financial support  under grant number
SAP105354. 
\section{Appendix}
\label{sec:app}
In this appendix we show how to obtain the infinite sums in Eq. (\ref{hggyu}).
Using the equations of motion for the fermion field profiles before the electroweak symmetry breaking, from the $S_{matter}$ one gets
\bea\nonumber
\partial_y{\hat\psi}_L + M_\psi(y){\hat\psi}_L = e^{A(y)}m_n{\hat\psi}_R,
\eea
\bea\nonumber
-\partial_y{\hat\psi}_R + M_\psi(y){\hat\psi}_R = e^{A(y)}m_n{\hat\psi}_L, 
\eea
 and using the definition (\ref{bm1}) we obtain
\bea
m_n{\hat\psi}_R-e^{-A-Q}\partial_y({\hat\psi}_Le^{Q})=0,
\eea
\bea
m_n{\hat\psi}_L+e^{-A+Q}\partial_y({\hat\psi}_Re^{-Q})=0,
\eea
where the hatted functions are defined as ${\hat\psi}\equiv e^{-2A}\psi$. 
We now multiply the first equation by $e^{A+Q}$ and the second by $e^{A-Q}$. Integrating from $0$ to some arbitrary value, $y^{\prime}$,
 gives,
\bea
\int_0^{y'}e^{A+Q}{\hat\psi}_R={1\over m_n}{\hat\psi}_L(y')e^{Q(y')},
\eea
\bea
\int_0^{y'}e^{A-Q}{\hat\psi}_L=-{1\over m_n}{\hat\psi}_R(y')e^{-Q(y')},
\eea
where we have imposed Dirichlet boundary conditions, ${\hat\psi}_L(0)=0$ on the first equation, and ${\hat\psi}_R(0)=0$ on the second one. 
Multiplying these equations by ${\hat\psi}_R(y'')$ and ${\hat\psi}_L(y'')$ and performing a summation over all of the KK modes we obtain
\bea
\int_0^{y'}e^{A+Q}\sum_{n=1}^{\infty}{\hat\psi}_R^{(n)}(y''){\hat\psi}_R^{(n)}(y)=e^{Q(y')}\sum_{n=1}^{\infty}{{\hat\psi}_R^{(n)}(y''){\hat\psi}_L^{(n)}(y')\over m_n},
\eea
\bea
\int_0^{y'}e^{A-Q}\sum_{n=1}^{\infty}{\hat\psi}_L^{(n)}(y''){\hat\psi}_L^{(n)}(y)=-e^{-Q(y')}\sum_{n=1}^{\infty}{{\hat\psi}_L^{(n)}(y''){\hat\psi}_R^{(n)}(y')\over m_n}.
\eea
Now using the completeness of the Sturm-Liouville system as\footnote{Note that here we are working with the hatted functions ${\hat\psi}\equiv e^{-2A}\psi$.}
\bea
\sum_{n=0}^{\infty}{\hat\psi}^{(n)}(y){\hat\psi}^{(n)}(y')=e^{-A}\delta(y-y'),
\eea
we obtain the sums
\bea
\sum_{n=1}^{\infty}{{\hat\psi}_R^{(n)}(y''){\hat\psi}_L^{(n)}(y')\over m_n}=e^{-Q(y')}\int_0^{y'}e^{A+Q}\left[e^{-A}\delta(y''-y)-{\hat\psi}^{(0)}_R(y''){\hat\psi}^{(0)}_R(y)\right],
\eea
\bea
\sum_{n=1}^{\infty}{{\hat\psi}_L^{(n)}(y''){\hat\psi}_R^{(n)}(y')\over m_n}=-e^{Q(y')}\int_0^{y'}e^{A-Q}\left[e^{-A}\delta(y''-y)-{\hat\psi}^{(0)}_L(y''){\hat\psi}^{(0)}_L(y)\right].
\eea
Finally performing the $\delta$-function integrals and using the normalized zero modes ($y_1$ being the position of the IR brane)
\bea
{\hat\psi}_{L}^{(0)}(y)={e^{-Q(y)}\over(\int_0^{y_1}e^{A-2Q})^{1\over2}}, \ {\hat\psi}_{R}^{(0)}(y)={e^{Q(y)}\over(\int_0^{y_1}e^{A+2Q})^{1\over2}},
\eea
we get
\bea\label{sumuf}
\sum_{n=1}^{\infty}{{\hat\psi}_R^{(n)}(y''){\hat\psi}_L^{(n)}(y')\over m_n}=e^{Q(y'')-Q(y')}\left[\theta(y'-y'')-{\int_0^{y'}e^{A-2Q}\over\int_0^{y_1}e^{A-2Q} }\right],
\eea
\bea\label{sumqf}
\sum_{n=1}^{\infty}{{\hat\psi}_L^{(n)}(y''){\hat\psi}_R^{(n)}(y')\over m_n}=-e^{Q(y')-Q(y'')}\left[\theta(y'-y'')-{\int_0^{y'}e^{A+2Q}\over\int_0^{y_1}e^{A+2Q} }\right].
\eea


\end{document}